\documentclass[runningheads]{svmult}
\usepackage{makeidx}   
\usepackage{epsfig}
\usepackage{rotate}
\usepackage{latexsym}
\usepackage{amsmath}
\usepackage{mathrsfs}
\usepackage{amsfonts}
\usepackage{graphicx}

\usepackage{subeqnar}  
\usepackage{multicol}  
\usepackage{cropmark} 
\usepackage{physprbb}  
\newcommand{\bea}{\begin{eqnarray}}
\newcommand{\eea}{\end{eqnarray}}
\newcommand{\be}{\begin{equation}}
\newcommand{\ee}{\end{equation}}

\begin{document}

\title*{Liouville Cosmology}
\toctitle{Liouville Cosmology}
%
%
\titlerunning{Liouville Cosmology}
%
\author{
\rightline{gr-qc/0502119}
\rightline{CERN-PH-TH/2005-037, ACT-02-05, MIFP-05-05}
~\\
~\\
John~Ellis$^1$
\and N.~E.~Mavromatos$^2$
\and D.~V.~Nanopoulos$^3$
}
\authorrunning{John Ellis et al.}
%
%
\institute{1 - TH Division, Physics Department, CERN\\
2 - Theoretical Physics, Physics Department, King's College 
London, UK\\
3 - Mitchell Institute for Fundamental Physics, Texas A\&M
University, College Station, TX 77843, USA;\\
Astroparticle Physics Group, Houston
Advanced Research Center,
Mitchell Campus,
Woodlands, TX~77381, USA;\\
Academy of Athens,
Academy of Athens,
Division of Natural Sciences, 28~Panepistimiou Avenue, Athens 10679,
Greece}

\maketitle              

\begin{abstract}
Liouville string theory is a natural framework for discussing the
non-equilibrium evolution of the Universe. It enables non-critical strings
to be treated in mathematically consistent manner, in which target time
is identified with a world-sheet renormalization-group scale parameter,
preserving target-space general coordinate invariance and the existence of
an S-matrix. We review our proposals for a unified treatment of inflation
and the current acceleration of the Universe. We link the current
acceleration of the Universe with the value of the string coupling. In
such a scenario, the dilaton plays an essential background r\^ole, driving
the acceleration of the Universe during the present era after decoupling
as a constant during inflation.
\end{abstract}


\section{Issues in String Cosmology} 

Formal developments in string theory~\cite{gsw} over the past
decade~\cite{polchinski}, with the discovery of a consistent way of
studying quantum domain-wall structures (D-branes), have opened up novel
ways of looking at not only the microcosmos, but also the macrocosmos.

In the microcosmos, there are novel ways of compactification, either via
the observation~\cite{add} that extra dimensions that are large compared
to the string scale~\cite{add} are consistent with the foundations of
string theory, or by viewing our four-dimensional world as a brane
embedded in a higher-dimensional bulk space-time, whose extra dimensions
might even be infinite in size~\cite{randal}. Such ideas are still
consistent with the required large hierarchy between the Planck scale and
the electroweak- or supersymmetry-breaking scale. In this modern approach,
fields in the gravitational (super)multiplet of the (super)string are
allowed to propagate in the bulk, but not the gauge fields, which are
attached to the brane world. In this way, the weakness of gravity compared
to the rest of the interactions is a result of the large compact
dimensions, and the compactification is not necessarily achieved through
the conventional means of closing the extra dimensions up on spatial
compact manifolds, but could also involve shadow brane worlds with special
reflecting properties, such as orientifolds, which restrict the bulk
dimension~\cite{ibanez}. In such approaches, the string scale $M_s$ is not
necessarily identical to the four-dimensional Planck mass $M_P$. Instead,
they are related by
\begin{equation}
M_P^2 = \frac{8M_s^2 V_6}{g_s^2} .
\label{planckstring}
\end{equation}
through the large compactification volume $V_6$.

In the macrocosmos, this modern approach has offered new insights into the
evolution of our Universe.  Novel ways of discussing cosmology in brane
worlds have been discovered over the past five years, which may
revolutionise our way of approaching issues such as
inflation~\cite{langlois,ekpyrotic}.  On the other hand, mounting
experimental evidence from diverse astrophysical sources presents some
important issues that string theory must address if it is to provide a
realistic description of Nature. Observations of distant
supernovae~\cite{snIa} and the cosmic microwave background fluctuations,
e.g., by the WMAP satellite~\cite{wmap}, indicate that the expansion of
our Universe is currently accelerating, and that 73\% of its energy
density consists of some unknown Dark Energy.

This cosmological development may be quite significant for string theory,
requiring that we revolutionise the approach usually followed so far. If
the dark energy turns out to be an honest-to-God cosmological constant,
leading to an asymptotic de Sitter horizon, then the entire concept of the
scattering matrix, the basis of perturbative string theory, breaks down.
This would cast doubts on the very foundations of string theory, at least
in its familiar formulation~\cite{EW}. Alternatively, one might
invoke some quintessential model for the vacuum energy, in which the
vacuum energy relaxes to zero at large cosmic time. This may be consistent
with the existence of an S-matrix as well as with the astrophysical data,
but there is still the open issue of embedding such models in
(perturbative) string theory. In particular, one must develop a consistent
$\sigma$-model formulation of strings propagating in such relaxing,
time-dependent space-time backgrounds.

The world-sheet conformal-invariance conditions of critical string theory,
which are equivalent to the target-space equations of motion for the
background fields on which the string propagates, are very restrictive,
corresponding only to vacuum solutions of critical strings. The main
problem may be expressed as follows.  Consider the graviton world-sheet
$\beta$ function, which is nothing but the Ricci tensor of the target
space-time background to lowest order in $\alpha '$:
\begin{equation} 
\beta_{\mu\nu} = \alpha ' R_{\mu\nu}~,
\label{beta} 
\end{equation}
in the absence of other fields.
Conformal invariance requires the vanishing of $\beta_{\mu\nu}$,
which implies that the background must be Ricci-flat, a characteristic
of solutions to the {\it vacuum} Einstein equations.

The issue then arises how to describe cosmological string backgrounds,
which are not vacuum solutions, but require the presence of a matter
fluid, yielding a non-flat Ricci tensor. Specifically, a vacuum
solution with a cosmological constant is inconsistent with conformal
invariance: a de Sitter Universe with a positive cosmological constant
$\Lambda > 0$ has a Ricci tensor $R_{\mu\nu} = \Lambda g_{\mu\nu}$, where
$g_{\mu\nu}$ is the metric tensor.

We now discuss how these issues may be addressed in Liouville Cosmology.

\section{Universes in Dilaton Backgrounds}

A proposal for obtaining a non-zero cosmological constant in string theory
was made in~\cite{fischler}, according to which dilaton tadpoles in
higher-genus world-sheet surfaces produce additional modular infinities.
Their regularisation would lead to extra world-sheet structures in the
$\sigma$ model that do not appear at the world-sheet tree level, leading
to modifications of the $\beta$-function. As a result, the Ricci tensor of
the space-time background is now that of a (anti) de Sitter Universe, with
a cosmological constant given by the dilaton tadpole graph $J_D >0$ ($J_D
< 0$). One problem with this approach is the above-mentioned existence of
an asymptotic horizon in the de Sitter case, which prevents the proper
definition of asymptotic states, and hence a scattering S-matrix.  
However, since the perturbative world-sheet formalism is based on such an
S-matrix, one may question the fundamental consistency of this approach.

A way out of this dilemma was proposed in~\cite{aben}, namely a dilaton
background depending linearly on time in the so-called $\sigma$-model
frame. Such a background, even when the $\sigma$-model metric is flat,
leads to solutions of the conformal invariance conditions of the pertinent
stringy $\sigma$-model that are exact to all orders in $\alpha'$, thereby
constituting acceptable solutions from a perturbative view point. It was
argued in~\cite{aben} that such backgrounds describe linearly-expanding
Robertson-Walker (RW) Universes, which were shown to be exact
conformal-invariant solutions, corresponding to Wess-Zumino models of
appropriate group manifolds.  The important novelty in~\cite{aben} was the
identification of target time $t$ with a specific dilaton background:
\begin{equation}
\label{lineardil}
\Phi = {\rm const} - Q~t,
\end{equation}
where $Q$ is a constant.

The square of $Q$ is the $\sigma$-model central charge deficit in a {\it
supercritical} string theory~\cite{aben}. Consistency of the underlying
world-sheet conformal field theory, as well as modular invariance of the
string scattering amplitudes, led to {\it discrete} values of $Q^2$, when
expressed in units of the string length $M_s$. This was actually the first
example of a non-critical string, with the target-space coordinates $X^i$,
$i=1, \dots D-1$ playing the r\^oles of the pertinent $\sigma$-model
fields. This non-criticality broke conformal invariance, which was
compensated by Liouville dressing~\cite{ddk}. The required Liouville field
had time-like signature, since the central-charge deficit was positive:
$Q^2 > 0$ in the model of~\cite{aben}, and played the r\^ole of target
time.

In the presence of a non-trivial dilaton field, the 
Einstein term in the effective $D$-dimensional low-energy
field theory action is conformally rescaled by $e^{-2\Phi}$.
This requires a redefinition 
of the $\sigma$-model-frame space-time metric $g_{\mu\nu}^\sigma$ to
become the physical metric in the `Einstein frame', $g_{\mu\nu}^E$:
\begin{equation}
g_{\mu\nu}^E = e^{\frac{2\Phi}{D-2}}g_{\mu\nu}^\sigma.
\label{smodeinst}
\end{equation}
A redefinition of target time is also necessary to obtain the
standard RW form of the metric in the
Einstein frame:
\begin{equation}
ds^2_E = -dt_E^2 + a_E^2(t_E) \left(dr^2 + r^2 d\Omega^2 \right),
\end{equation}
where we assume for definiteness a spatially-flat RW metric. Here
$a_E$ is an appropriate scale factor, which is a function of $t_E$ alone
in the homogeneous cosmological backgrounds we assume throughout.
Time in the Einstein-frame is related to time in the $\sigma$-model 
frame~\cite{aben} by:
\begin{equation}
\label{einsttime}
dt_E = e^{-\Phi}dt \qquad \to \qquad t_E = \int ^t e^{-\Phi(t)} dt~. 
\end{equation} 
The linear dilaton background (\ref{lineardil}) therefore yields
the following relation between the Einstein and $\sigma$-model
time variables:
\begin{equation} 
t_E = c_1 + \frac{c_0}{Q}e^{Qt},
\end{equation}
where $c_{1,0}$ are appropriate (positive) constants.

Thus, the dilaton background (\ref{lineardil})
scales logarithmically with 
the Einstein-Robertson-Walker cosmic time $t_E$:
\begin{equation}\label{dil2}
\Phi (t_E) =({\rm const.}') - {\rm ln}(\frac{Q}{c_0}t_E)
\end{equation} 
In this regime, the string coupling, which is defined as~\cite{gsw}: 
\begin{equation}
g_s = {\rm exp}\left(\Phi(t)\right)
\label{defstringcoupl}
\end{equation}
varies with the cosmic time $t_E$ as
\begin{equation}
g_s^2 (t_E) \equiv e^{2\Phi} \propto \frac{1}{t_E^2}.
\label{vanishes}
\end{equation}
Thus, the effective string coupling vanishes
asymptotically in cosmic time. 


The effective low-energy action for the gravitational multiplet 
of the string in the Einstein frame reads~\cite{aben}:
\begin{equation}
S_{\rm eff}^{\rm brane} = \int d^4x\sqrt{-g}\{ R  - 2(\partial_\mu \Phi)^2 
- \frac{1}{2} e^{4\Phi}( \partial_\mu b)^2 - \frac{1}{3}e^{2\Phi}\delta c 
\},
\label{effaction}
\end{equation}
\noindent
where $b$ is the four-dimensional axion field
associated with the antisymmetric tensor and $\delta c = C_{\rm int} -
c^*$ is the central charge deficit, where $C_{\rm int}$ is the central 
charge
of the conformal world-sheet theory corresponding to the transverse
(internal) string dimensions, and $c^*=22 (6)$ is the critical value of
this internal central charge of the (super)string theory in a flat
four-dimensional space-time.  The linear-dilaton configuration
(\ref{dil2}) corresponds, in this language, to a background charge $Q$ of
the conformal theory, which contributes a term $-3Q^2$ (in our
normalisation (\ref{dil2})) to the total central charge, which also
receives contributions from the four uncompactified dimensions of our
world.  In the case of a flat four-dimensional Minkowski space-time, one
has:  $C_{\rm total} = 4 - 3Q^2 + C_{\rm int} = 4 - 3Q^2 + c^* + \delta 
c$, which should equal 26. This implies that $C_{\rm int} = 22 + 3Q^2~(6 +
3Q^2)$ for bosonic (supersymmetric) strings.  

An important result
in~\cite{aben} was the discovery of an exact conformal field theory
corresponding to the dilaton background (\ref{dil2}), i.e., a constant
curvature (Milne) (static) metric in the $\sigma$-model frame or,
equivalently, a linearly-expanding RW Universe in the
Einstein frame.  This conformal field theory corresponds to a
two-dimensional Wess-Zumino-Witten (WZW) model on the world sheet, on a
group manifold $O(3)$ with appropriate constant curvature, whose
coordinates correspond to the spatial components of the four-dimensional
metric and antisymmetric tensor fields, together with a free world-sheet
field corresponding to the target time coordinate. The total central
charge in this more general case reads $C_{\rm total} = 4 -3Q^2 -
\frac{6}{k+2}+ C_{\rm int}$ with $k$ a positive integer, which corresponds
to the level of the Kac-Moody algebra associated with the WZW model on the
group manifold. The value of $Q$ is chosen in such a way that the overall
central charge $c=26$ and the theory is conformally invariant.

It was observed in~\cite{aben} that known unitary conformal field theories
have {\it discrete} central charges, which accumulate to integers or
half-integers from {\it below}, and hence that the values of the central
charge deficit $\delta c$ are also {\it discrete}. From a physical point
of view, this implies that the linear-dilaton Universe may either stay at
such a state for ever with a fixed $\delta c$, or else tunnel through
different discrete levels before relaxing to a critical $\delta c =0$
theory corresponding to a flat four-dimensional Minkowski space-time.

The analysis in~\cite{aben} also showed that there
were tachyonic mass-squared shifts of order $-Q^2$ for the bosonic string
excitations, but not for the fermionic ones. This in turn would imply the
breaking of target supersymmetry in such backgrounds, as far as the
excitation spectrum is concerned, and the appearance of tachyonic
instabilities. The latter could trigger these cosmological phase
transitions, since they correspond to world-sheet operators that are
relevant in the renormalization-group sense. As such, they can trigger the
flow of the internal unitary conformal field theory towards minimisation
of its central charge, according to the Zamolodchikov
$c$ theorem~\cite{zam}. As we discuss below, in semi-realistic
cosmological models~\cite{dgmpp}, such tachyons decouple from the spectrum
relatively quickly. On the other hand, as a result of the form of the
dilaton in the Einstein frame (\ref{dil2}), we observe that the
dark-energy density for this Universe, $\Lambda \equiv e^{2\Phi}\delta c$,
relaxes towards zero as $1/t_E^2$, for each of the stationary values of
$\delta c$.  The breaking of supersymmetry induced by the linear dilaton
should therefore be considered an {\it obstruction}~\cite{witten}, rather
than a spontaneous breaking, in the sense of appearing only in the
boson-fermion mass splittings between the excitations, whereas the vacuum
energy of the asymptotic equilibrium theory vanishes.

In~\cite{emn} we went one step further than~\cite{aben}, considering more
complicated $\sigma$-model metric backgrounds in $(D+1)$-dimensional
target space-times, that did not satisfy the $\sigma$-model
conformal-invariance conditions, and therefore were in need of Liouville
dressing~\cite{ddk}. These backgrounds were even allowed to be
time-dependent. Non-criticality can be introduced in many mathematically
consistent ways, for instance via cosmically catastrophic events such as
the collision of brane worlds~\cite{gravanis,brany}, which lead naturally
to supercritical $\sigma$ models. The Liouville dressing of such
non-critical models results in $D+2$-dimensional target spaces with two
time directions.  The important point of~\cite{emn} was the {\it
identification} of the world-sheet zero mode of the Liouville field with
the target time, thereby restricting the Liouville-dressed $\sigma$ model
to a $(D+1)$-dimensional hypersurface of the $(D+2)$-dimensional target
space-time, maintaining the initial target space-time dimensionality. We
stress once more that this identification is only possible in cases where
the initial $\sigma$ model is supercritical, so that the Liouville mode
has time-like signature~\cite{aben,ddk}.  Such an identification was shown
in certain models~\cite{gravanis,brany} to be energetically
preferable from a target-space viewpoint, since it minimised certain
effective potentials in the low-energy field theory corresponding to the
string theory at hand.

Such non-critical $\sigma$-models relax asymptotically in the cosmic
Liouville time to conformal $\sigma$ models, which may be viewed as
equilibrium points in string theory space. In some interesting cases of
relevance to cosmology~\cite{dgmpp}, which were particularly generic, the
asymptotic conformal field theory was that of~\cite{aben} with a linear
dilaton and a flat Minkowski target-space metric in the $\sigma$-model
frame. In others, the asymptotic theory was characterised by a constant
dilaton and a Minkowski space-time~\cite{gravanis}.  In what follows, we
describe briefly the main features of such non-critical cosmological
string models, and compare them with recent observations.

\section{Non-Critical Liouville String Cosmologies} 

We now consider in more detail the model of~\cite{dgmpp}. Although
formulated in the specific framework of ten-dimensional
Type-0~\cite{type0} string theory. This has a non-supersymmetric
target-space spectrum, thanks to a special projection of the
supersymmetric partners out of the spectrum. Nevertheless, its basic
cosmological properties are sufficiently generic to be extended to the
bosonic sector of any effective low-energy supersymmetric field theory
obtained from a supersymmetric string model.

The ten-dimensional metric configuration considered in~\cite{dgmpp} 
was: 
\begin{equation}
G_{MN}=\left(\begin{array}{ccc}g^{(4)}_{\mu\nu} \qquad 0 \qquad 0 \\
0 \qquad e^{2\sigma_1} \qquad 0 \\ 0 \qquad 0 \qquad
e^{2\sigma_2} I_{5\times 5} \end{array}\right),
\label{metriccomp}
\end{equation}
where lower-case Greek indices are four-dimensional space-time
indices, and $I_{5\times 5}$ denotes the $5\times 5$ unit matrix.
We have chosen two different scales for internal space. The field
$\sigma_{1}$ sets the scale of the fifth dimension, while
$\sigma_{2}$ parametrises a flat five-dimensional space. In the
context of the cosmological models we deal with here, the
fields $g_{\mu\nu}^{(4)}$, $\sigma_{i},~i=1,2$ are assumed to
depend only on the time $t$.

Type-0 string theory, as well as its supersymmetric extensions that
appear, e.g., in brane models, contains form fields with non-trivial gauge
fluxes (flux-form fields), which live in the higher-dimensional bulk
space. In the specific model of~\cite{type0}, there is one such field that
was assumed to be non-trivial. As was demonstrated in~\cite{dgmpp}, a
consistent background choice for the flux-form field has a flux parallel
to to the fifth dimension $\sigma_2$. This implies that the internal space
is crystallised (stabilised)  in such a way that this dimension is much
larger than the remaining four, demonstrating the physical importance of
the flux fields for large radii of compactification.

Considering the fields to be time-dependent only, i.e., considering
spherically-symmetric homogeneous backgrounds, restricting attention to
the compactification (\ref{metriccomp}), and assuming a RW
form of the four-dimensional metric, with scale factor $a(t)$, the
generalised conformal invariance conditions and the Curci-Pafutti
$\sigma$-model renormalisability constraint~\cite{curci} imply a set of
differential equations which were solved numerically in \cite{dgmpp}. The
generic form of these equations reads~\cite{ddk,emn,dgmpp}:
\begin{equation} 
  {\ddot g}^i + Q(t){\dot g}^i = -{\tilde \beta}^i ,
\label{liouvilleeq}
\end{equation} 
where the ${\tilde \beta}^i$ are the Weyl anomaly coefficients of the 
stringy $\sigma$ model on the background $\{ g^i \}$. 
In the model of~\cite{dgmpp}, the set of $\{ g^i \}$ includes the
graviton, dilaton, tachyon, flux and moduli fields $\sigma_{1,2}$, 
whose vacuum expectation values control the size of the extra dimensions.

The detailed analysis of\cite{dgmpp} indicated that the moduli $\sigma_i$
fields froze quickly to their equilibrium values. Thus, they together with
the tachyon field, which also decays rapidly to a constant value, decouple
from the four-dimensional fields at very early stages in the evolution of
this string Universe~\footnote{The presence of the tachyonic instability
in the spectrum is due to the fact that in Type-0 strings there is no
target-space supersymmetry by construction.  From a cosmological
viewpoint, the tachyon fields are not necessarily bad features, since they
may provide the initial instability leading to cosmic
expansion~\cite{dgmpp}, as well as a mechanism for step-wise reduction in
the central-charge deficit.}. There is an inflationary phase in this
scenario and a dynamical exit from it. The important point that guarantees
the exit is the fact that the central-charge deficit $Q^2$ is a
time-dependent entity in this approach, obeying specific relaxation laws
determined by the underlying conformal field
theory~\cite{dgmpp,gravanis,brany}. The central charge runs with the local
world-sheet renormalisation-group scale, namely the zero mode of the
Liouville field, which is identified~\cite{emn} with the target time in
the $\sigma$-model frame.  The supercriticality~\cite{aben} $Q^2 > 0$ of
the underlying $\sigma$ model is crucial, as already mentioned.  
Physically, the non-critical string provides a model of
non-equilibrium dynamics, which may be the result of some catastrophic
cosmic event, such as a collision of two brane
worlds~\cite{ekpyrotic,gravanis,brany}, or an initial quantum
fluctuation~\cite{emninfl,dgmpp}.  It also provides, as we now discuss
briefly, a unified mathematical framework for analysing various phases of
string cosmology, including the inflationary phase in the early Universe,
graceful exit from it and reheating, as well as the current and future
eras of accelerated cosmologies. It is interesting that one can constrain
string parameters such as string coupling, the separation of brany 
worlds at the end of
inflation and the recoil velocity of the branes after the collision, by
fits to current astrophysical data~\cite{brany}.

\section{Liouville Inflation: the Big Picture} 

As discussed in~\cite{emninfl,gravanis,brany}, 
a constant central-charge deficit
$Q^2$ in a stringy $\sigma$ model may be associated with an initial 
inflationary phase, with
\begin{equation}
\label{centraldeficit} 
Q^2 = 9 H^2 > 0~,
\end{equation}
where the Hubble parameter $H$ may be fixed in terms of other parameters
of the model. One can consider various scenarios for such a departure from
criticality. For example, in the specific colliding-brane model
of~\cite{gravanis,brany}, $Q$ (and thus $H$) is proportional to the square
of the relative velocity of the colliding branes, $Q \propto u^2$ during
the inflationary era.  As is evident from (\ref{centraldeficit}) and
discussed in more detail below, in a phase of constant $Q$ one obtains an
inflationary de Sitter Universe.

The specific normalization in (\ref{centraldeficit}) is imposed by the
identification of the time $t$ with (minus) the zero mode of the Liouville
field $-\varphi$ of the {\it supercritical } $\sigma$ model. The minus
sign may be understood both mathematically, as due to properties of the
Liouville mode, and physically by the requirement that the deformation of
the space-time relaxes following the distortion induced by the recoil.
With this identification, the general equation of motion for the couplings
$\{ g_i \}$ of the $\sigma$-model background modes is~\cite{emn}:
\begin{equation}
{\ddot g}^i + Q{\dot g}^i 
= -\beta^i (g) = -{\cal G}^{ij} \partial C[g]/\partial g^j~,
\label{liouveq}
\end{equation}
where the dot denotes a derivative with respect to the Liouville
world-sheet zero mode $\varphi$, and ${\cal G}^{ij}$ is an inverse
Zamolodchikov metric in the space of string theory couplings $\{ g^i
\}$~\cite{zam}. When applied to scalar inflaton-like string modes,
(\ref{liouveq})  would yield standard field equations for scalar fields in
de Sitter (inflationary)  space-times, provided the normalization
(\ref{centraldeficit}) is valid, implying a `Hubble' expansion parameter
$H=-Q/3$~\footnote{The gradient-flow property of the $\beta$ functions
makes the analogy with the inflationary case even more profound, with the
running central charge $C[g]$~\cite{zam} playing the r\^ole of the
inflaton potential in conventional inflationary field theory.}. The minus
sign in $Q=-3H$ is due to the sign in the identification of the target 
time $t$ with the world-sheet zero mode of $-\varphi$~\cite{emn}.

The relations (\ref{liouveq}) generalize and replace the
conformal-invariance conditions $\beta^i = 0$ of the critical string
theory, and express the conditions necessary for the restoration of
conformal invariance by the Liouville mode~\cite{ddk}. Interpreting the
latter as an extra target dimension, the conditions (\ref{liouveq}) may
also be viewed as conformal invariance conditions of a {\it critical}
$\sigma$ model in (D+1) target space-time dimensions, where D is the
target dimension of the non-critical $\sigma$ model before Liouville
dressing.  In most Liouville approaches, one treats the Liouville mode
$\varphi$ and time $t$ as independent coordinates.  However, in our
approach~\cite{emn,dgmpp,gravanis}, as already mentioned, we take the
further step of restricting this extended (D+1)-dimensional space-time to
a hypersurface determined by the identification $\varphi = -t$. This means
that, as time flows, one is restricted to a D-dimensional subspace of the
full (D+1)-dimensional Liouville space-time. This restriction arose in the
work of~\cite{gravanis,brany} because the potential between massive
particles in the effective field theory was found to be proportional to
${\rm cosh}(t + \varphi)$, which is minimized when $\varphi = -t$.

However, the flow of the Liouville mode opposite to that of target time
may be given a deeper mathematical interpretation.  It may be viewed as a
consequence of a specific treatment of the area constraint in non-critical
(Liouville) $\sigma$ models~\cite{emn}, which involves the evaluation of
the Liouville-mode path integral via an appropriate steepest-descent
contour.  In this way, one obtains a `breathing' world-sheet evolution, in
which the world-sheet area starts from a very large value (serving as an
infrared cutoff), shrinks to a very small one (serving as an ultraviolet
cutoff), and then inflates again towards very large values (returning to
an infrared cutoff). Such a situation may then be interpreted~\cite{emn}
as a world-sheet `bounce' back to the infrared, implying that the physical
flow of target time is opposite to that of the world-sheet scale
(Liouville zero mode).

We now become more specific. We consider a non-critical $\sigma$ model
with a background metric $G_{\mu\nu}$, antisymmetric tensor $B_{\mu\nu}$,
and dilaton $\Phi$. These have the following ${\cal O}(\alpha')$ $\beta$
functions, where $\alpha'$ is the Regge slope~\cite{gsw}:
\begin{eqnarray} 
&& \beta^G_{\mu\nu} = \alpha ' \left( R_{\mu\nu} + 2 \nabla_{\mu} 
\partial_{\nu} \Phi 
 - \frac{1}{4}H_{\mu\rho\sigma}H_{\nu}^{\rho\sigma}\right)~, \nonumber \\
&& \beta^B_{\mu\nu} = \alpha '\left(-\frac{1}{2}\nabla_{\rho} H^{\rho}_{\mu\nu} + 
H^{\rho}_{\mu\nu}\partial_{\rho} \Phi \right)~, \nonumber \\
&& {\tilde \beta}^\Phi = \beta^\Phi - \frac{1}{4}G^{\rho\sigma}\beta^G_{\rho\sigma} = 
\frac{1}{6}\left( C - 26 \right).
\label{bfunctions}
\end{eqnarray} 
The Greek indices are four-dimensional, including target-time
components $\mu, \nu, ...= 0,1,2,3$ on the D3-branes
of~\cite{gravanis}, and $H_{\mu\nu\rho}= \partial_{[\mu}B_{\nu\rho]}$ is the
antisymmetric tensor field strength.
We consider the following representation of the four-dimensional 
field strength in terms of a pseudoscalar (axion-like) field $b$: 
\begin{equation}
H_{\mu\nu\rho} = \epsilon_{\mu\nu\rho\sigma}\partial^\sigma b ,
\label{axion}
\end{equation}
where $\epsilon_{\mu\nu\rho\sigma}$ is the four-dimensional antisymmetric
symbol. Next, we choose an axion background that is linear in the
time $t$~\cite{aben}:
\begin{equation} 
b = b(t) = \beta t~, \quad  \beta={\rm constant} ,
\label{axion2}
\end{equation}
which yields a constant field strength with spatial indices only: $H_{ijk}
= \epsilon_{ijk}\beta$, $H_{0jk}= 0$.  This implies that such a background
is a conformal solution of the full ${\cal O}(\alpha')$ $\beta$ function
for the four-dimensional antisymmetric tensor. We also consider a dilaton
background that is linear in the time $t$~\cite{aben}:
\begin{equation}
\Phi (t,X) = {\rm const} + ({\rm const})' t .
\label{constdil}
\end{equation}
This background does not contribute to the $\beta$ functions 
for the antisymmetric and metric tensors.

Suppose now that only the metric is a non-conformal background, due to 
some initial quantum fluctuation or catastrophic event, such as the 
collision of two branes discussed above, 
which results in an initial central charge deficit $Q^2$ 
(\ref{centraldeficit}) that is constant at early stages after the 
collision. Let 
\begin{equation} 
G_{ij} = e^{\kappa \varphi + Hct}\eta_{ij}~, \quad G_{00}=e^{\kappa '\varphi 
+ Hct}\eta_{00},
\label{metricinfl}
\end{equation}
where $t$ is the target time, $\varphi$ is the Liouville mode, 
$\eta_{\mu\nu}$ is the four-dimensional Minkowski metric, 
and $\kappa, \kappa'$ and $c$ are constants to be determined. 
As already discussed, the standard inflationary scenario in 
four-dimensional physics requires $Q = -3H$,
which stems from the identification
of the Liouville mode with time~\cite{emn}
$\varphi = -t$, that is imposed dynamically~\cite{gravanis}
at the end of our computations. Initially, one should treat
$\varphi, t$ as independent target-space components. 

The Liouville dressing induces~\cite{ddk} $\sigma$-model terms of the form 
$\int_{\Sigma} R^{(2)} Q \varphi$, where $R^{(2)}$ is the world-sheet curvature.
Such terms provide non-trivial contributions to the dilaton background in 
the (D+1)-dimensional space-time $(\varphi,t,X^i)$:
\begin{equation}
\Phi (\varphi,t,X^i) = Q \,\varphi + ({\rm const})' t + {\rm const}.
\label{seventy}
\end{equation}
If we choose 
\begin{equation}
({\rm const})'=Q~, 
\label{const=Q}
\end{equation}
we see that (\ref{seventy}) implies a 
{\it constant} dilaton background during the inflationary era, in which the 
central charge deficit $Q$ is constant. 

The choices (\ref{seventy}) and (\ref{const=Q}), like the identification
$\varphi = -t$, apply to the world-sheet zero modes of the Liouville field
and the time coordinate. As such, they imply a constant dilaton at the
mean-field (classical) level. World-sheet quantum fluctuations of the
dilaton, associated with non-zero modes of these fields, do not cancel,
since the identification $\varphi = -t$ is not valid for the fluctuating
parts of the respective $\sigma$-model fields. This leads in turn to
non-trivial fluctuations of the dilaton field during inflation.  The
summation over world-sheet genera turns such fluctuations into 
target-space
quantum fluctuations. This allows one~\cite{brany} to apply the
phenomenology of scalar field fluctuations used in conventional
inflationary models also in this case, in order to constrain physically
important parameters of the non-critical string theory by means of recent
cosmological data~\cite{wmap}.

We now consider the Liouville-dressing equations~\cite{ddk}
(\ref{liouveq}) for the $\beta$ functions of the metric and antisymmetric
tensor fields (\ref{bfunctions}) at the level of 
world-sheet zero modes of the $\sigma$-model fields. The
dilaton equation yields no independent information for a constant 
mean dilaton field, apart from expressing
the dilaton $\beta$ function in terms of the central-charge deficit as
usual. For the axion background (\ref{axion2}), only the metric yields a
non-trivial constraint (we work in units with $\alpha' =1$ for
convenience):
\begin{equation} 
{\ddot G}_{ij} + Q{\dot G}_{ij} = -R_{ij} + \frac{1}{2}\beta^2 G_{ij},
\end{equation}
where the dot indicates differentiation with respect to the 
(zero mode of the) world-sheet Liouville mode $\varphi$, and $R_{ij}$ is 
the
(non-vanishing) Ricci tensor of the (non-critical) $\sigma$ model with
coordinates $(t,{\vec x})$:  $R_{00}=0~, R_{ij}=\frac{c^2H^2}{2}e^{(\kappa
- \kappa ')\varphi}\eta_{ij}$. One should also take into account the
temporal equation for the metric tensor:
\begin{equation}
{\ddot G}_{00} + Q{\dot G}_{00} = -R_{00} = 0,
\label{tempgrav}
\end{equation}
where the vanishing of the Ricci tensor stems from the 
specific form of the background (\ref{metricinfl}). The analogue 
equation is identically zero for the antisymmetric tensor background.
We seek metric backgrounds of inflationary (de Sitter) RW 
form:
\begin{equation}
G_{00}=-1~, \quad G_{ij}=e^{2Ht}\eta_{ij}.
\label{desittermetric}
\end{equation}
Then, from (\ref{desittermetric}), (\ref{metricinfl}),
(\ref{constdil}) and (\ref{axion2}), we observe that there indeed is a 
consistent solution with:
\begin{equation}
Q = -3H = - \kappa ',~c=3,~\kappa = H,~\beta^2 = 5H^2,
\label{solution}
\end{equation}
corresponding to the conventional form of inflationary equations for
scalar fields.

In this talk we do not mention ways of exiting from this inflationary
phase and reheating the Universe. These issues may also be approached from
a Liouville $\sigma$-model point of view, as we shall report in a
forthcoming publication~\cite{EMNW}.

\section{Liouville Dark Energy: the End Game}

In the generic class of non-critical string models described in this talk,
the $\sigma$ model always asymptotes, for long enough cosmic times, to the
linear-dilaton conformal $\sigma$-model field theory of`\cite{aben}.
However, it is important to stress that this is only an asymptotic limit,
and the current era of our Universe should be viewed as close to, but
still not at the equilibrium relaxation point. Thus the dilaton is almost
linear in the $\sigma$-model time, and hence varies almost logarithmically
in the Einstein-frame time (\ref{dil2}). This slight non-equilibrium would
lead to a time dependence of the unified gauge coupling and other
constants such as the four-dimensional Planck length (\ref{planckstring}),
mainly through the time-dependence of the string coupling
(\ref{defstringcoupl}) that results from the time dependence of the linear
dilaton (\ref{lineardil}).

The asymptotic-time regime of the Type-0 cosmological string model
of~\cite{dgmpp} has been obtained analytically, by solving the pertinent
equations (\ref{liouvilleeq}) for the various fields. As already
mentioned, at later times the theory becomes four-dimensional, and the
only non-trivial information is contained in the scale factor and the
dilaton, given that the topological flux field remains conformal in this
approach, and the moduli and initial tachyon fields decouple very fast at
the initial stages after inflation in this model. For times long after the
initial fluctuations, such as the present times where the linear
approximation is valid, the solution for the dilaton in the $\sigma$-model
frame, as follows from the equations (\ref{liouvilleeq}), takes the form:
\begin{equation} \label{dilaton} 
\Phi (t) =-{\rm ln}\left[\frac{\alpha A}{F_1}{\rm cosh}(F_1t)\right],
\end{equation}
where $F_1$ is a positive constant, $\alpha$ is a numerical constant of 
order one, and 
\begin{equation}\label{defA2}
A = \frac{C_5 e^{s_{01}}}{\sqrt{2}V_6}~, 
\end{equation}
with $s_{01}$ the equilibrium
value of the modulus field $\sigma_1$ associated with the large bulk 
dimension, and $C_5$ the corresponding flux of the five-form flux field.
Notice the independence of $A$ from this large bulk dimension.

For very large times $F_1 t \gg 1$ (in string units), one therefore
approaches a linear solution for the dilaton: $\Phi \sim {\rm const} -F_1
t$. From (\ref{dilaton}), (\ref{defstringcoupl}) and (\ref{planckstring}),
we thus observe that the asymptotic weakness of gravity in this
Universe~\cite{dgmpp} is due to the smallness of the internal space $V_6$
as compared with the flux $C_5$ of the five-form field. The constant $F_1$
is related to the central-charge deficit of the underlying non-conformal
$\sigma$-model~\cite{dgmpp}:
\begin{equation}\label{ccd}
Q = q_0 + \frac{q_0}{F_1}(F_1 + \frac{d\Phi}{dt}),
\end{equation}
where $q_0$ is a constant, and the numerical solution of 
(\ref{liouvilleeq}) studied in \cite{dgmpp})
requires that $q_0/F_1 =(1 + \sqrt{17})/2 \simeq 2.56$.
However, we believe that this is only a result of the numerical 
approximations in the analysis of \cite{dgmpp}, 
and for our purposes we consider from now on 
\begin{equation}
F_1 \sim q_0,
\label{f1q0}
\end{equation}
in accord with~\cite{aben}, to which the model relaxes for large times. In
this spirit, we require that the value of $q_0$ to which the central
charge deficit (\ref{ccd}) asymptotes must be, for the consistency of the
underlying string theory, one of the discrete values obtained
in~\cite{aben}, for which the string scattering amplitudes factorise. This
asymptotic string theory with a time-independent central-charge deficit,
$q_0^2 \propto c^*-25 $ (or $c^*-9$ for superstring)  may therefore be
considered as an {\it equilibrium} situation, with an $S$-matrix defined
for specific (discrete) values of the central charge $c^*$, generalizing
the standard critical (super)string which corresponds to central charge
$c^*=25$ (=9 for superstrings)~\cite{ddk,aben}.

Defining the Einstein frame time $t_E$ through (\ref{einsttime}),
we obtain in this case (\ref{dilaton}) 
\begin{equation}\label{einstframe}
t_E=\frac{\alpha A}{F_1^2}sinh(F_1 t).
\end{equation}
In terms of the Einstein-frame time 
one obtains a logarithmic time-dependence~\cite{aben} for the dilaton
\begin{equation} 
\Phi _E = {\rm const} -{\rm ln}(\gamma t_E)~,
\label{einsteindil} 
\end{equation}
where
\begin{equation}\label{defA} 
\gamma \equiv \frac{F_1^2}{\alpha A}~. 
\end{equation}
For large $t_E$, e.g., now or in the future, one has
\begin{equation}\label{einstmetr} 
a_E(t_E) \simeq \frac{F_1}{\gamma}\sqrt{1 + \gamma^2 t_E^2}.
\end{equation}
At very large times $a(t_E)$ scales linearly with the Einstein-frame
cosmological time $t_E$~\cite{dgmpp}, and hence there is no cosmic
horizon. From a field-theoretical viewpoint, this would allow for a proper
definition of asymptotic states and thus a scattering matrix.  As we
mentioned briefly above, however, from a stringy point of view, there are
restrictions in the asymptotic values of the central charge deficit $q_0$,
and there is only a discrete spectrum of values of $q_0$ that allow for a
full stringy S-matrix to be defined, respecting modular
invariance~\cite{aben}. Asymptotically, the Universe relaxes to its
ground-state equilibrium situation, and the non-criticality of the string
caused by the initial fluctuation disappears, giving rise to a critical
(equilibrium) string Universe with a Minkowski metric and a linear-dilaton
background.  This is a generic feature of the models considered here and
in~\cite{emn04}, allowing the conclusions to be extended beyond Type-0
string theory to incorporate also string/brane models with target-space
supersymmetry, such as those in~\cite{emw,brany}.

An important comment is in order at this point, regarding the 
form of the Einstein metric corresponding to (\ref{einstmetr}):
\begin{equation}\label{einstmetr2}
g_{00}^E=-1, ~~\quad g_{ij} = a_E^2(t_E) 
= \frac{F_1^2}{\gamma^2} + F_1^2t_E^2~.
\end{equation}
Although asymptotically for $t_E \to \infty$ the above metric asymptotes
to the linearly-expanding Universe of \cite{aben}, the presence of a
constant $F_1^2/\gamma^2$ contribution implies that the solution for large
but finite $t_E$, such as the current era of the Universe, is different
from that of~\cite{aben}.  Indeed, the corresponding $\sigma$-model metric
(\ref{smodeinst}) is not Minkoswski-flat, and the pertinent
$\sigma$ model does not correspond to a conformal field theory. This
should come as no surprise, because for finite $t_E$, no matter how large,
the $\sigma$-model theory requires Liouville dressing. It is only at the
end-point of the time-flow: $t_E \to \infty$ that the underlying string
theory becomes conformal, and the system reaches equilibrium.

The Hubble parameter of such a Universe becomes for large $t_E$ 
\begin{equation}\label{hubble2} 
H(t_E) \simeq \frac{\gamma^2 t_E}{1 + \gamma^2 t_E^2}~.
\end{equation}
On the other hand, the Einstein-frame effective four-dimensional 
`vacuum energy density', defined through the running 
central-charge deficit $Q^2$,
upon compactification to four dimensions of the ten-dimensional
expression $\int d^{10}x \sqrt{-g}e^{-2\Phi}Q^2(t_E)$, is~\cite{dgmpp}:
\begin{equation} 
\Lambda_E (t_E) = e^{2\Phi - \sigma_1 - 5\sigma_2}Q^2(t_E) 
\simeq \frac{q_0^2 \gamma^2}{F_1^2 ( 1 + \gamma^2 t_E^2)}
\sim \frac{\gamma^2}{1 + \gamma^2 t_E^2},
\label{cosmoconst2} 
\end{equation}
where, for large $t_E$, $Q$ is given in (\ref{ccd}) and approaches its
equilibrium value $q_0$, and we took into account (\ref{f1q0}). Thus, the
dark energy density relaxes to zero for $t_E \to \infty$. Notice an
important feature of the form of the relaxation (\ref{cosmoconst2}),
namely that the proportionality constants in front are such that, for
asymptotically large $t_E \to \infty$, $\Lambda$ becomes
independent of the equilibrium conformal field theory central charge
$q_0$.

Finally, and most importantly for our purposes here, 
the deceleration parameter in the same regime of $t_E$ becomes:
\begin{equation} 
q(t_E) = -\frac{(d^2a_E/dt_E^2)~a_E}{({da_E/dt_E})^2} 
\simeq -\frac{1}{\gamma^2 t_E^2}.
\label{decel4}
\end{equation}
The important point to make in connection 
with this expression is that, as is clear from 
(\ref{einsteindil}) and (\ref{defstringcoupl}),
it can be identified, up
to irrelevant constants of proportionality which conventional
normalisation sets to one, 
with the square of the string coupling~\cite{emn04}:
\begin{equation}
|q(t_E)| = g_s^2
\label{important}
\end{equation}
To guarantee the consistency of perturbation theory, one must have $g_s < 
1$, which can be achieved in our approach if one defines the  
present era by the time regime
\begin{equation}
\gamma \sim t_E^{-1} 
\label{condition}
\end{equation} 
in the Einstein frame.
This is compatible with large enough times $t_E$ (in string units) 
for 
\begin{equation} 
|C_5|e^{-5s_{02}}/F_1^2 \sim |C_5|e^{-5s_{02}}/q_0^2 \gg 1~,
\label{largetwe}
\end{equation} 
as becomes clear from (\ref{defA2}),(\ref{defA}) and (\ref{f1q0}).  This
condition can be guaranteed either by small radii of five of the extra
dimensions, or by a large value of the flux $|C_5|$ of the five-form of
the Type-$0$ string, compared with $q_0$. We discuss later concrete
examples of non-critical string cosmologies, in which the asymptotic value
of the central charge $q_0 \ll 1$ in string units.  Recalling that the
relatively large extra dimension in the direction of the flux, $s_{01}$,
decouples from this condition, we observe that there is the
possibility of constructing effective five-dimensional models with a large
uncompactified fifth dimension while respecting the condition
(\ref{condition}).

The Hubble parameter and the cosmological constant continue to be
compatible with the current observations in the regime (\ref{condition})
of Einstein-frame times, while the string coupling (\ref{important}) is
kept finite and of order unity by the conditions (\ref{decel4},
\ref{condition}), as suggested by grand unification
phenomenology~\cite{gsw}.

\section{Dark Energy and the String Coupling}

We next turn to the equation of state of our Universe.
As discussed in~\cite{dgmpp}, our model resembles
quintessence models, with the dilaton playing the r\^ole of the 
quintessence field. Hence the equation of state
for our Type-$0$ string Universe reads~\cite{carroll}:
\begin{equation}\label{eqnstate} 
          w_\Phi = \frac{p_\Phi}{\rho_\Phi}=\frac{\frac{1}{2}({\dot \Phi})^2 - V(\Phi)}
{\frac{1}{2}({\dot \Phi})^2 + V(\Phi)},
\end{equation}
where $p_\Phi$ is the pressure and $\rho_\Phi$ is the energy density, and
$V(\Phi)$ is the effective potential for the dilaton, which in our case is
provided by the central-charge deficit term.  Here the dot denotes
Einstein-frame differentiation.  In the Einstein frame, the potential
$V(\Phi)$ is given by $\Lambda_E $ in (\ref{cosmoconst}). In the limit $Q
\to q_0$, which we have argued should characterise the present era to a
good approximation, the effective potential $V(\Phi)$ is then of order
$(q_0^2/2F_1^2)t_E^{-2}$, where we recall (c.f., (\ref{f1q0}))  that
$q_0/F_1$ is of order one. In the Einstein frame the exact normalisation of 
the dilaton field is $\Phi _E = {\rm const} -{\rm ln}(\gamma t_E) $.
We then obtain for the present era: 
\begin{equation}\label{dilpotkin}  
\frac{1}{2}{\dot \Phi}^2 \sim \frac{1}{2t_E^2}, \qquad V(\Phi) 
\sim \frac{6.56}{2}\frac{1}{t_E^2}.
\end{equation} 
This implies an equation of state (\ref{eqnstate}): 
\begin{equation} 
w_\Phi (t_E \gg 1) \simeq -0.74 
\label{eqnstatedil}
\end{equation}
for (large) times $t_E$ in string units corresponding to the present era
(\ref{condition}). This number can be pushed lower, towards $w \to -1$, 
by a slight adjustment of the various parameters, improving the agreement 
with current cosmological data~\cite{wmap}.
Assuming a conventional effective four-dimensional low-energy 
fluid Universe, which is a good picture in our situation 
where the moduli and other fields have decoupled at early
stages of the Universe, we have:
\begin{equation} 
q = \frac{1}{2}(1 + 3w_\Phi)
\label{wq}
\end{equation}
from which we obtain 
\begin{equation}
q = - 0.61.
\label{valueq}
\end{equation}
This fixes the string coupling (\ref{important}) in the perturbative 
regime
consistent with grand unification scenarios extrapolated from low 
energies.

So far the model does not include ordinary matter: only fields from
the string gravitational multiplet have been included.
Inclusion of ordinary matter is not expected to change
qualitatively the result. We conjecture that the 
fundamental relation (\ref{important}) will continue to hold,
the only difference being that the inclusion of 
ordinary matter will tend to reduce the string acceleration:
\begin{equation}
q = \frac{1}{2}\Omega_M - \Omega_\Lambda~,
\label{mattercosmo}
\end{equation}
where $\Omega_M (\Omega_\Lambda)$ denotes the total matter
(vacuum) energy density, normalised to the critical energy density 
of a spatially flat Universe.

{\it There is a remarkable coincidence in numbers for this 
non-supersymmetric
Type-0 string Universe with the astrophysical observations}, which yield
$q$ close to th value (\ref{valueq}). The ordinary matter content of the
Universe has $\Omega_{\rm ordinary~matter} \simeq 0.04$ and the dark
matter content is estimated to have $\Omega_{DM} = 0.23$, while the Dark
Energy content is $\Omega_\Lambda \simeq 0.73$. This yields $q = -0.595$,
which is only a few per cent away from (\ref{valueq}). In fact, if one
naively used the value (\ref{valueq}) for $q$, obtained in our case where
ordinary matter was ignored, in the expression (\ref{mattercosmo}), one
would find $\Omega_\Lambda \simeq 0.74$, indicating that the contribution
of the dilaton field to the cosmic acceleration is the dominant one.
  
If the relation (\ref{important})  holds after the inclusion of matter,
even in supersymmetric models, one arrives at an even better value of the
string coupling, $g^2_s \simeq 0.595$, more consistent with the 
unification
prediction of the minimal supersymmetric standard model at a scale $\sim
10^{16}$ GeV. The only requirement for the asymptotic condition
(\ref{important}) to hold is that the underlying stringy $\sigma$ model is
non-critical and asymptotes for large times to the linear dilaton
conformal field theory of~\cite{aben}. It should be understood, of course,
that the precise relation of the four-dimensional gauge coupling with the
ten-dimensional string coupling depends on the details of
compactification, which we do not discuss here.

The variation of the dilaton field at late cosmic times implies a slow
variation of the string coupling (\ref{important}), ${\dot g_s}/g_s =
1/t_E \sim 10^{-60}$ in the present era. The corresponding variations of
the gauge couplings are too small to affect current phenomenology.

The above considerations are rather generic to models which relax
asymptotically to the linear-dilaton conformal field theory solutions
of~\cite{aben}, and from this point of view are physically interesting. We
did not need to specify above the microscopic theory underlying the
deviation from non criticality.  For this one would need some specific
example of such a deviation from a conformally-invariant point in string
theory space. One such example, with physically interesting consequences,
is provided by the colliding brane-world scenario, in which the Liouville
string $\sigma$ model describes stringy excitations on the brane worlds,
at relatively long times after the collision so that string perturbation
theory is valid. We now discuss this briefly.

\section{A Concrete Non-Critical String Example: Colliding Branes}

\begin{figure}[htb]
\begin{center}
\epsfxsize=3in
\bigskip
\centerline{\epsffile{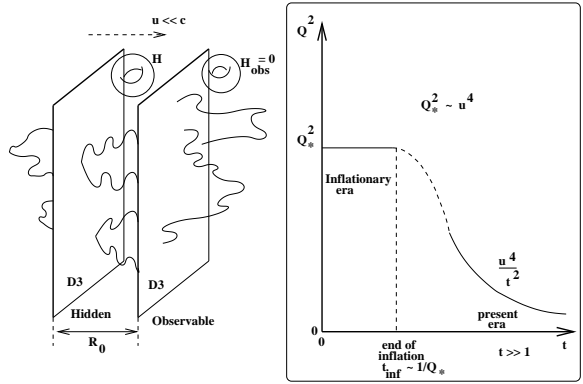}}
\caption{{\it A scenario 
in which the collision of two Type-II 5-branes provides 
inflation and a relaxation model for cosmological vacuum energy.\label{infla}}}
\end{center} 
\end{figure}

We now concentrate on particular examples of the previous general
scenario~\cite{emninfl}, in which the non-criticality is induced by the
collision of two branes as seen in Fig.~\ref{infla}. We first discuss the
basic features of this scenario. For our purposes below we assume that the
string scale is of the same order as the four-dimensional Planck scale,
though this is an assumption that may be relaxed, in view of recent
developments in strings with large compactification directions, as we
mentioned in the Introduction.

Following~\cite{gravanis}, we consider two 5-branes of Type-II string
theory, in which the extra two dimensions have been compactified on tori.
On one of the branes (assumed to be the hidden world), the torus is
magnetized with a field intensity ${\cal H}$.  Initially our world is
compactified on a normal torus, without a magnetic field, and the two
branes are assumed to be on a collision course with a small relative
velocity $v \ll 1$ in the bulk, as illustrated in Fig.~\ref{infla}. The
collision produces a non-equilibrium situation, which results in electric
current transfer from the hidden brane to the visible one.  This causes
the (adiabatic) emergence of a magnetic field in our world.

The instabilities associated with such magnetized-tori compactifications
are not a problem in the context of the cosmological scenario discussed
here. As discussed in~\cite{gravanis}, the collision may also produce
decompactification of the extra toroidal dimensions at a rate much slower
than any other rate in the problem. As also discussed in~\cite{gravanis},
this guarantees asymptotic equilibrium and a proper definition of an
$S$-matrix for the stringy excitations on the observable world. We shall
come back to this issue at the end of this Section.

\paragraph{}
\begin{figure}[tb]
\begin{center}
\includegraphics[width=4cm]{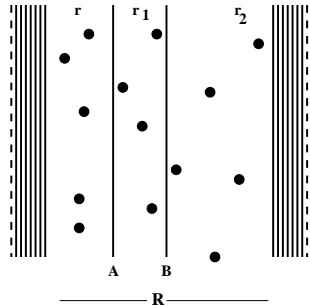}
\end{center}
\caption{\it A model for supersymmetric D-particle foam
consisting of two stacks each of sixteen parallel coincident D8-branes, 
with orientifold planes (thick dashed lines) attached to them~\cite{emw}.
The space does not extend beyond the orientifold planes.
The bulk region of ten-dimensional space in which the D8-branes 
are embedded is punctured by D0-branes (D-particles, dark blobs). 
The two parallel stacks are sufficiently far from each other
that any Casimir contribution to the vacuum energy is negligible.
If the branes are stationary, there is zero vacuum energy, and the 
configuration 
is a consistent supersymmetric string vacuum. To obtain excitations
corresponding to interesting cosmologies, one should move one (or more) of 
the branes from each 
stack and then let them collide (Big Bang), bounce back (inflation),
and then relax to their original position, where they collide again
with the remaining branes in each stack (exit from inflation, reheating).}
\label{fig:nonchiral}
\end{figure}

The collision of the two branes implies, for a short period afterwards,
while the branes are at most a few string scales apart, the exchange of
open-string excitations stretching between the branes, whose ends are
attached to them. As argued in~\cite{gravanis}, the exchanges of such
pairs of open strings in Type-II string theory result in an excitation
energy in the visible world.  The latter may be estimated by computing the
corresponding scattering amplitude of the two branes, using string-theory
world-sheet methods~\cite{bachas}: the time integral for the relevant
potential yields the scattering amplitude.  Such estimates involve the
computation of appropriate world-sheet annulus diagrams, due to the
existence of open string pairs in Type-II string theory. This implies the
presence of `spin factors' as proportionality constants in the scattering
amplitudes, which are expressed in terms of Jacobi $\Theta$ functions. For
the small brane velocities $v \ll 1$ we are considering here, the
appropriate spin structures start at quartic order in $v$, for the case of
identical branes, as a result of the mathematical properties of the Jacobi
functions~\cite{bachas}. This in turn implies~\cite{gravanis,emw} that the
resulting excitation energy on the brane world is of order $V = {\cal
O}(v^4)$, which may be thought of as an initial (approximately constant)
value of a supercritical central-charge deficit for the non-critical
$\sigma$ model that describes stringy excitations in the observable world
after the collision:
\begin{equation}
Q^2 = \left(\sqrt{\beta} v^2  + {\cal H}^2\right)^2 > 0.
\label{initialdeficit}
\end{equation} 
where, in the model of \cite{emw,brany}, 
the proportionality factor $\beta$, 
computed using string amplitude computations,
is of order
\begin{equation} 
\beta \sim 2\sqrt{3} \cdot 10^{-8} \cdot g_s~, 
\label{betaorder}
\end{equation} 
with $g_s$ the string coupling $g_s^2 \sim 0.5$ for interesting 
phenomenological models~\cite{gsw,ibanez}, as discussed above. 
We recall that the supercriticality, i.e., the positive 
definiteness of the central charge deficit (\ref{initialdeficit}),  
of the model is essential~\cite{aben} 
for a time-like signature of the Liouville mode, and hence its 
interpretation as target time.

At times long after the collision, the branes slow down and the central 
charge deficit is no
longer constant but relaxes with time $t$.  In the approach
of~\cite{gravanis}, this relaxation has been computed by using world-sheet
logarithmic conformal field theory methods~\cite{kogan}, taking into
account recoil (in the bulk) of the observable-world brane and the
identification of target time with the (zero mode of the) Liouville field.
In that work it was assumed that the final equilibrium 
value of the central charge deficit was zero, i.e., the theory approached
a critical string. This late-time varying deficit $Q^2(t)$ 
scales with the target time (Liouville mode) as (in units 
of the string scale $M_s$):
\begin{equation} 
Q^2 (t) \sim \frac{({\cal H}^2 + v^2)^2}{t^2}.
\label{cosmoconst}
\end{equation} 
Some explanations are necessary at this point.
In arriving at (\ref{cosmoconst}), one identifies 
the world-sheet renormalisation group scale ${\cal T} ={\rm ln}(L/a)^2$
(where $(L/a)^2$ is the world-sheet area), appearing in the Zamolodchikov 
$c$-theorem used to determine the rate of change of $Q$,  
with the zero mode of a normalised 
Liouville field $\phi_0$, such that $\phi_0 = Q{\cal T}$. This normalisation
guarantees a canonical kinetic term for the Liouville field in the 
world-sheet action~\cite{ddk}. Thus it is $\phi_0$ that is identified
with $-t$, with $t$ the target time.

On the other hand, in other models~\cite{dgmpp} the asymptotic
value of the central-charge 
deficit may not be zero, in the sense that the asymptotic theory 
is that of a linear dilaton, with a Minkowski metric in the 
$\sigma$-model frame~\cite{aben}. This theory is still a conformal
model, but the central charge is a constant $q_0$, and in fact the 
dilaton is of the form $\Phi = q_0 t + {\rm const}$, 
where $t$ is the target time in the $\sigma$-model frame.
Conformal invariance, as mentioned previously, requires~\cite{aben}
that $q_0$ takes on 
one of a discrete set of values, in the way explained
in~\cite{aben}.
In such a case, following the same method as in the $q_0=0$ case 
of~\cite{gravanis}, one arrives at the asymptotic form
\begin{equation} 
Q^2 (t) \sim q_0^2 + {\cal O}\left(\frac{{\cal H}^2 + v^2)}{t}q_0\right) 
\label{cosmoconst3}
\end{equation} 
for large times $t$. 

The colliding-brane model of \cite{gravanis} can be extended to
incorporate proper supersymmetric vacuum configurations of string
theory~\cite{emw}. As illustrated in Fig.~\ref{fig:nonchiral}, this model
consists of two stacks of D8-branes with the same tension, separated by a
distance $R$. The transverse bulk space is restricted to lie between two
orientifold planes, and is populated by D-particles.  It was shown
in~\cite{emw} that, in the limit of static branes and D-particles, this
configuration constitutes a zero vacuum-energy supersymmetric ground state
of this brane theory. Bulk motion of either the D-branes or the
D-particles~\footnote{The latter could arise from recoil following
scattering with closed-string states propagating in the bulk.} results in
non-zero vacuum energy~\cite{emw} and hence the breaking of target
supersymmetry, proportional to some power of the average (recoil) velocity
squared, which depends on the precise string model used to described the
(open) stringy matter excitations on the branes.

The colliding-brane scenario can be realized~\cite{brany} in this
framework by allowing (at least one of) the D-branes to move, keeping the
orientifold planes static. One may envisage a situation in which the two
branes collide at a certain moment in time corresponding to the Big Bang -
a catastrophic cosmological event setting the beginning of observable time
- and then bounce back. The width of the bulk region is assumed to be long
enough that, after a sufficiently long time following the collision, the
excitation energy on the observable brane world - which corresponds to the
conformal charge deficit in a $\sigma$-model framework~\cite{gravanis,emw}
- relaxes to tiny values. It is expected that a ground-state configuration
will be achieved when the branes reach the orientifold planes again
(within stringy length uncertainties of order $\ell_s=1/M_s$, the string
scale). In this picture, since observable time starts ticking after the
collision, the question how the brane worlds started to move is merely
philosophical or metaphysical. The collision results in a kind of phase
transition, during which the system passes through a non-equilibrium
phase, in which one loses the conformal symmetry of the stringy $\sigma$
model that describes perturbatively string excitations on the branes. At
long times after the collision, the central charge deficit relaxes to
zero~\cite{gravanis}, indicating that the system approaches equilibrium
again. The Dark Energy observed today may be the result of the fact that
our world has not yet relaxed to this equilibrium value. Since the
asymptotic ground state configuration has static D-branes and D-particles,
and hence has zero vacuum energy as guaranteed by the exact conformal
field theory construction of~\cite{emw,brany}, it avoids the fine tuning
problems in the model of~\cite{gravanis}.

Sub-asymptotically, there are several contributions to the excitation
energy of our brane world in this picture. One comes from the interaction
of the brane world with nearby D-particles, i.e., those within distances
at most of order ${\cal O}(\ell_s)$, as a result of open strings stretched
between them.  The other contribution comes from the collision of the
identical D-branes.  For a sufficiently dilute gas of nearby D-particles
we may assume that this latter contribution is the dominant one. In this
case, one may ignore the D-particle/D-brane contributions to the vacuum
energy, and hence apply the previous considerations on inflation, based on
the ${\cal O}(v^4)$ central charge deficit, with $v$ the velocity of the
brane world in the bulk.

The presence of D-particles, which inevitably cross the D-branes in such a
picture, even if the D-particle defects are static initially, distorts
slightly the inflationary metric on the observable brane world at early
times after the collision, during an era of approximately constant central
charge deficit, without leading to significant qualitative changes.
Moreover, the existence of D-particles on the branes will affect the
propagation of string matter on the branes, in the sense of modifying
their dispersion relations by inducing local curvature in space-time, as a
result of recoil following collisions with string matter. However, it was
argued in~\cite{emnequiv} that only photons are susceptible to such
effects in this scenario, due to the specific gauge properties of the
membrane theory at hand.  The dispersion relations for chiral matter
particles, or in general fields on the D-branes that transform
non-trivially under the Standard Model gauge group, are protected by
special gauge symmetries in string theory, and as such are not modified.

\section{Liouville's Dark Secrets} 

The use of Liouville strings to describe the evolution of our Universe
seems generally appropriate, since non-critical strings are associated
with non-equilibrium situations which undoubtedly occurred in the Early
Universe, and may still occur today.  It is remarkable that the departure
from criticality may even enhance the predictability of string theory,
although the space of non-critical string theories is much larger that of
critical strings, to the point that purely stringy quantities such as the
string coupling are accessible to experiment.

We have discussed in this talk Liouville cosmological models based on
non-critical strings with various asymptotic configurations of the
dilaton, speculating on the Big Bang itself, on the inflationary phase and
the possibility of exit from it, as well the evolution of the Universe at
large times, both current and future. A particularly interesting case from
the physical point of view is that of a dilaton that is asymptotically
linear in cosmic time. This is known to correspond to a proper conformal
field theory~\cite{aben}.  We have observed that the string coupling is
identified in such a model (up to irrelevant constants of order
one)~\cite{emn04} with the deceleration parameter of the Universe through
equation (\ref{important}).

We stress once more the importance of non-criticality in arriving at
(\ref{important}). In critical strings, which usually assume the absence
of a four-dimensional dilaton, such a relation is not obtained, and the
string coupling is not directly measurable in this way.

The approach of the identification of target time in such a framework with
a world-sheet renormalisation group scale, the Liouville mode~\cite{emn},
provides a novel way of selecting the ground state of the string theory,
which may not necessarily be associated with minimisation of energy, but
could be a matter of cosmic `chance'. Indeed, it may be a random event
that the initial state of our cosmos corresponds to a certain Gaussian
fixed point in the space of string theories, which is then perturbed in
the Big Bang by some relevant (in a world-sheet sense) deformation,
thereby making the theory non-critical, and hence out of equilibrium from
a target space-time viewpoint. Then the theory flows along some
renormalisation-group trajectory to some specific ground state,
corresponding to the infrared fixed point of this perturbed world sheet
$\sigma$-model theory. This approach allows for many parallel universes to
be implemented of course, and our world might be just one of these. Each
Universe, may flow between different fixed points, perturbed by different
operators. Standard world-sheet renormalization-group arguments imply that
the various flow trajectories do not intersect, although this is something
that is far from proven in general. It seems to us that this scenario is
much more specific than the landscape scenario~\cite{sussk}, which has
recently been advocated as an attempt to parametrise our ignorance of the
true structure of string/M theory.

\section*{Acknowledgements}

This work is based on a talk given by D.V.N. at the conference
DARK 2004 in Texas A \& M University in October 2004.  
N.E.M. wishes to thank Juan Fuster and IFIC-University of Valencia 
(Spain) for their interest and support. 
The work of D.V.N. is supported by D.O.E. grant
DE-FG03-95-ER-40917.

\end{document}